# Diversity-Multiplexing Tradeoff in Selective-Fading Multiple-Access MIMO Channels

Pedro Coronel, Markus Gärtner and Helmut Bölcskei

Communication Technology Laboratory
ETH Zurich, 8092 Zurich, Switzerland
E-mail: {pco, gaertner, boelcskei}@nari.ee.ethz.ch

*Abstract*— We establish the optimal diversity-multiplexing (DM) tradeoff of coherent selective-fading multiple-access multiple-input multiple-output (MIMO) channels and provide corresponding code design criteria. As a byproduct, on the conceptual level, we find an interesting relation between the DM tradeoff framework and the notion of dominant error event regions which was first introduced in the AWGN case by Gallager, *IEEE Trans. IT*, 1985. This relation allows to accurately characterize the error mechanisms in MIMO fading multiple-access channels. In particular, we find that, for a given rate tuple, the maximum achievable diversity order is determined by the error event that dominates the total error probability exponentially in SNR. Finally, we show that the distributed space-time code construction proposed recently by Badr and Belfiore, *Int. Zurich Seminar on Commun.*, 2008, satisfies the code design criteria derived in this paper.

## I. INTRODUCTION

The diversity-multiplexing (DM) tradeoff framework introduced by Zheng and Tse allows to efficiently characterize the information-theoretic performance limits of communication over multiple-input multiple-output (MIMO) fading channels both in the point-to-point [1] and in the multiple-access (MA) case [2]. For coherent[1] point-to-point flat-fading channels, DM tradeoff optimal code constructions have been reported in [3]–[6]. The optimal DM tradeoff in point-to-point selective-fading MIMO channels was recently characterized in [7]. In the MA case, the optimal DM tradeoff is known only for flat-fading channels [2]. Corresponding DM tradeoff optimal code constructions were recently reported in [8], [9].

*Contributions:* The aim of this paper is to characterize the DM tradeoff in selective-fading MIMO multiple-access channels (MACs) and to derive corresponding code design criteria. As a byproduct, on the conceptual level, we find an interesting relation between the DM tradeoff framework and the notion of error event regions which was first introduced in the AWGN case by Gallager in [10] and recently applied to MIMO fading MACs in [11]. This relation leads to an accurate characterization of the error mechanisms in MIMO fading MACs. Furthermore, we extend the techniques introduced in [7] for computing the DM tradeoff in point-to-point selective-fading channels to the MA case. Finally, we prove that the distributed space-time block codes proposed in [9] satisfy the code design criteria derived in this paper.

*Notation:* $M_T$ and $M_R$ denote, respectively, the number of transmit antennas for each user and the number of receive antennas. The set of all users is $\mathcal{U} = \{1, 2, \ldots, U\}$, $\mathcal{S}$ is a subset of $\mathcal{U}$ with $\bar{\mathcal{S}}$ and $|\mathcal{S}|$ denoting its complement in $\mathcal{U}$ and its cardinality, respectively. The superscripts $^T$ and $^H$ stand for transposition and conjugate transposition, respectively. $\mathbf{A} \otimes \mathbf{B}$ and $\mathbf{A} \odot \mathbf{B}$ denote, respectively, the Kronecker and Hadamard products of the matrices $\mathbf{A}$ and $\mathbf{B}$. If $\mathbf{A}$ has the columns $\mathbf{a}_k$ ($k = 1, 2, \ldots, m$), $\text{vec}(\mathbf{A}) = [\mathbf{a}_1^T\ \mathbf{a}_2^T\ \ldots\ \mathbf{a}_m^T]^T$. $\mathbf{A}^{1/2}$ denotes the positive semidefinite square root of $\mathbf{A}$. For index sets $\mathcal{I}_1 \subseteq \{1, 2, \ldots, n\}$ and $\mathcal{I}_2 \subseteq \{1, 2, \ldots, m\}$, $\mathbf{A}(\mathcal{I}_1, \mathcal{I}_2)$ stands for the (sub)matrix consisting of the rows of $\mathbf{A}$ indexed by $\mathcal{I}_1$ and the columns of $\mathbf{A}$ indexed by $\mathcal{I}_2$. The nonzero eigenvalues of the $n \times n$ Hermitian matrix $\mathbf{A}$, sorted in ascending order, are denoted by $\lambda_k(\mathbf{A})$, $k = 1, 2, \ldots, \text{rank}(\mathbf{A})$. The Kronecker delta function is defined as $\delta_{n,m} = 1$ for $n = m$ and zero otherwise. If $X$ and $Y$ are random variables (RVs), $X \sim Y$ denotes equivalence in distribution. $\mathbb{E}_X$ is the expectation operator with respect to (w.r.t.) the RV $X$. The random vector $\mathbf{x} \sim \mathcal{CN}(\mathbf{0}, \mathbf{C})$ is multivariate circularly symmetric complex Gaussian with $\mathbb{E}\{\mathbf{x}\mathbf{x}^H\} = \mathbf{C}$. The functions $f(x)$ and $g(x)$ are said to be exponentially equal, denoted by $f(x) \doteq g(x)$, if $\lim_{x \to \infty} \frac{\log f(x)}{\log x} = \lim_{x \to \infty} \frac{\log g(x)}{\log x}$. Exponential inequality, denoted by $\dot{\geq}$ and $\dot{\leq}$, is defined analogously.

## II. CHANNEL AND SIGNAL MODEL

We consider a selective-fading MAC where $U$ users, with $M_T$ transmit antennas each, communicate with a single receiver with $M_R$ antennas. The corresponding input-output relation is given by

$$\mathbf{y}_n = \sqrt{\frac{\mathsf{SNR}}{M_T}} \sum_{u=1}^{U} \mathbf{H}_{u,n}\, \mathbf{x}_{u,n} + \mathbf{z}_n,\ n = 0, 1, \ldots, N-1, \quad (1)$$

where the index $n$ corresponds to a time, frequency or time-frequency slot and SNR denotes the per-user signal-to-noise ratio at each receive antenna. The vectors $\mathbf{y}_n$, $\mathbf{x}_{u,n}$ and $\mathbf{z}_n$ denote, respectively, the $M_R \times 1$ receive signal vector, the $M_T \times 1$ transmit signal vector corresponding to the $u$th user, and the $M_R \times 1$ circularly symmetric complex Gaussian noise vector

---

This work was supported in part by the Swiss National Science Foundation (SNF) under grant No. 200020-109619 and by the STREP project No. IST-026905 MASCOT within the Sixth Framework Programme of the European Commission.

[1]Throughout the paper, we shall consider the coherent case, where the receiver has perfect channel state information (CSI) and the transmitter does not have CSI, but is aware of the channel law.

satisfying $\mathbb{E}\{\mathbf{z}_n \mathbf{z}_{n'}^H\} = \delta_{n,n'} \mathbf{I}_{\mathrm{M_R}}$, all for the $n$th slot. We assume that the receiver has perfect knowledge of all channels and the transmitters do not have CSI but are aware of the channel law.

We restrict our analysis to spatially uncorrelated Rayleigh fading channels so that, for a given $n$, $\mathbf{H}_{u,n}$ has i.i.d. $\mathcal{CN}(0,1)$ entries. The channels corresponding to different users are assumed to be statistically independent. We do, however, allow for correlation across $n$ for a given $u$, and assume, for simplicity, that all scalar subchannels have the same correlation function so that, in summary, $\mathbb{E}\{\mathbf{H}_{u,n}(i,j)\,(\mathbf{H}_{u',n'}(i',j'))^*\} = \mathbf{R}_{\mathbb{H}}(n,n')\,\delta_{u,u'}\,\delta_{i,i'}\,\delta_{j,j'}$, for $i, i' = 1, 2, \ldots, \mathrm{M_R}$, $j, j' = 1, 2, \ldots, \mathrm{M_T}$, and $n, n' = 0, 1, \ldots, N-1$. The covariance matrix $\mathbf{R}_{\mathbb{H}}$ is obtained from the channel's time-frequency correlation function [12]. In the sequel, we let $\rho \triangleq \mathrm{rank}(\mathbf{R}_{\mathbb{H}})$. For any set $\mathcal{S} = \{u_1, u_2, \ldots, u_{|\mathcal{S}|}\}$, we stack the corresponding users' channel matrices for a given slot index $n$ according to

$$\mathbf{H}_{\mathcal{S},n} = [\mathbf{H}_{u_1,n}\ \mathbf{H}_{u_2,n}\ \ldots\ \mathbf{H}_{u_{|\mathcal{S}|},n}]. \quad (2)$$

With this notation, it follows that

$$\mathbb{E}\{\mathrm{vec}(\mathbf{H}_{\mathcal{S},n})\,(\mathrm{vec}(\mathbf{H}_{\mathcal{S},n'}))^H\} = \mathbf{R}_{\mathbb{H}}(n,n')\,\mathbf{I}_{|\mathcal{S}|\mathrm{M_T M_R}}. \quad (3)$$

### III. Preliminaries

Assuming that all users employ i.i.d. Gaussian codebooks[2], the set of achievable rate tuples $(R_1, R_2, \ldots, R_U)$ for a given channel realization $\{\mathbf{H}_{u,n}\}$ is given by

$$\mathcal{R} = \left\{(R_1, R_2, \ldots, R_U) : \forall \mathcal{S} \subseteq \mathcal{U}, \right.$$
$$\left. R(\mathcal{S}) \le \frac{1}{N} \sum_{n=0}^{N-1} \log\det\left(\mathbf{I} + \frac{\mathsf{SNR}}{\mathrm{M_T}}\,\mathbf{H}_{\mathcal{S},n}\mathbf{H}_{\mathcal{S},n}^H\right)\right\} \quad (4)$$

where $R(\mathcal{S}) = \sum_{u \in \mathcal{S}} R_u$. If a given rate tuple $(R_1, R_2, \ldots, R_U) \notin \mathcal{R}$, we say that the channel is in outage w.r.t. this rate tuple. Denoting the corresponding outage event as $\mathcal{O}$, we have

$$\mathbb{P}(\mathcal{O}) = \mathbb{P}\left(\bigcup_{\mathcal{S} \subseteq \mathcal{U}} \mathcal{O}_{\mathcal{S}}\right) \quad (5)$$

where the $\mathcal{S}$-outage event $\mathcal{O}_{\mathcal{S}}$ is defined as

$$\mathcal{O}_{\mathcal{S}} = \left\{\{\mathbf{H}_{\mathcal{S},n}\}_{n=0}^{N-1} : \right.$$
$$\left. \frac{1}{N} \sum_{n=0}^{N-1} \log\det\left(\mathbf{I} + \frac{\mathsf{SNR}}{\mathrm{M_T}}\,\mathbf{H}_{\mathcal{S},n}\mathbf{H}_{\mathcal{S},n}^H\right) < R(\mathcal{S})\right\}. \quad (6)$$

Our goal is to characterize (5) as a function of the rate tuple $(R_1, R_2, \ldots, R_U)$ in the high-SNR regime and to find criteria on the users' codebooks guaranteeing that the corresponding error probability behaves exponentially in SNR like $\mathbb{P}(\mathcal{O})$. To this end, we shall employ the DM tradeoff framework [1], which, in its MA version [2], will be briefly summarized next.

---

[2]A standard argument along the lines of that used to obtain [1, Eq. 9] shows that this assumption does not entail a loss of optimality in the high SNR regime, relevant to the DM tradeoff.

In the DM tradeoff framework, the data rate of user $u$ scales with SNR as $R_u(\mathsf{SNR}) = r_u \log \mathsf{SNR}$, where $r_u$ denotes the multiplexing rate. Consequently, a sequence of codebooks $\mathcal{C}_{r_u}(\mathsf{SNR})$, one for each SNR, is required. We say that this sequence of codebooks constitutes a family of codes $\mathcal{C}_{r_u}$ operating at multiplexing rate $r_u$. The family $\mathcal{C}_{r_u}$ is assumed to have block length $N$. At any given SNR, $\mathcal{C}_{r_u}(\mathsf{SNR})$ contains codewords $\mathbf{X}_u = [\mathbf{x}_{u,0}\ \mathbf{x}_{u,1}\ \ldots\ \mathbf{x}_{u,N-1}]$ satisfying the per-user power constraint

$$\mathrm{Tr}(\mathbf{X}_u \mathbf{X}_u^H) \le \mathrm{M_T} N,\ \forall \mathbf{X}_u \in \mathcal{C}_{r_u},\ u = 1, 2, \ldots, U. \quad (7)$$

Since we are dealing with a MAC, the overall family of codes is given by $\mathcal{C}_{\boldsymbol{r}} = \mathcal{C}_{r_1} \times \mathcal{C}_{r_2} \times \cdots \times \mathcal{C}_{r_U}$, where $\boldsymbol{r} = (r_1, r_2, \ldots, r_U)$ denotes the multiplexing rate tuple. At a given SNR, the corresponding codebook $\mathcal{C}_{\boldsymbol{r}}(\mathsf{SNR})$ contains $\mathsf{SNR}^{Nr(\mathcal{U})}$ codewords with $r(\mathcal{U}) = \sum_{u=1}^U r_u$.

The DM tradeoff realized by $\mathcal{C}_{\boldsymbol{r}}$ is characterized by the function

$$d(\mathcal{C}_{\boldsymbol{r}}) = -\lim_{\mathsf{SNR}\to\infty} \frac{\log P_e(\mathcal{C}_{\boldsymbol{r}})}{\log \mathsf{SNR}}$$

where $P_e(\mathcal{C}_{\boldsymbol{r}})$ is the *total* error probability (that is, the probability for the receiver to make a detection error for at least one user) obtained through maximum-likelihood (ML) detection. The optimal DM tradeoff curve $d^\star(\boldsymbol{r}) = \sup_{\mathcal{C}_{\boldsymbol{r}}} d(\mathcal{C}_{\boldsymbol{r}})$, where the supremum is taken over all possible families of codes satisfying the power constraint (7), quantifies the maximum achievable diversity gain as a function of the multiplexing rate tuple $\boldsymbol{r}$ [1]. Since the outage probability $\mathbb{P}(\mathcal{O})$ is a lower bound (exponentially in SNR) on the error probability of any coding scheme [2, Lemma 7], we have

$$d^\star(\boldsymbol{r}) \le -\lim_{\mathsf{SNR}\to\infty} \frac{\log \mathbb{P}(\mathcal{O})}{\log \mathsf{SNR}} \quad (8)$$

where the outage event, defined in (5) and (6), is w.r.t. the rates $R_u(\mathsf{SNR}) = r_u \log \mathsf{SNR}$, $\forall u$. As an extension of the result for the flat-fading case [2], we shall show in this paper that (8) holds with equality also for selective-fading MACs. However, just like in the case of point-to-point channels, a direct characterization of the right-hand side (RHS) of (8) for the selective-fading case seems analytically intractable since one has to deal with the sum of correlated (recall that the $\mathbf{H}_{u,n}$ are correlated across $n$) terms in (6). In the next section, we show how the technique introduced in [7] for characterizing the DM tradeoff of point-to-point selective-fading MIMO channels can be extended to the MA case.

### IV. Computing the optimal DM tradeoff curve

#### A. Lower bound on $\mathbb{P}(\mathcal{O}_{\mathcal{S}})$

First, we derive a lower bound on the individual terms $\mathbb{P}(\mathcal{O}_{\mathcal{S}})$ that will be key in establishing the optimal DM tradeoff. We start by noting that for any set $\mathcal{S} \subseteq \mathcal{U}$, Jensen's inequality provides the following upper bound:

$$\frac{1}{N} \sum_{n=0}^{N-1} \log\det\left(\mathbf{I} + \frac{\mathsf{SNR}}{\mathrm{M_T}}\mathbf{H}_{\mathcal{S},n}\mathbf{H}_{\mathcal{S},n}^H\right)$$
$$\le \log\det\left(\mathbf{I} + \frac{\mathsf{SNR}}{\mathrm{M_T}N}\boldsymbol{\mathcal{H}}_{\mathcal{S}}\boldsymbol{\mathcal{H}}_{\mathcal{S}}^H\right) \triangleq \mathrm{J}_{\mathcal{S}} \quad (9)$$

where the "Jensen channel" [7] is defined as

$$\mathcal{H}_{\mathcal{S}} = \begin{cases} [\mathbf{H}_{\mathcal{S},0} \ \mathbf{H}_{\mathcal{S},1} \ \ldots \ \mathbf{H}_{\mathcal{S},N-1}], & \text{if } M_R \leq |\mathcal{S}|M_T, \\ [\mathbf{H}_{\mathcal{S},0}^H \ \mathbf{H}_{\mathcal{S},1}^H \ \ldots \ \mathbf{H}_{\mathcal{S},N-1}^H], & \text{if } M_R > |\mathcal{S}|M_T. \end{cases} \quad (10)$$

Consequently, $\mathcal{H}_{\mathcal{S}}$ has dimension $\text{m}(\mathcal{S}) \times N\text{M}(\mathcal{S})$ with $\text{m}(\mathcal{S}) \triangleq \min(|\mathcal{S}|M_T, M_R)$ and $\text{M}(\mathcal{S}) \triangleq \max(|\mathcal{S}|M_T, M_R)$. In the following, we say that the event $\mathcal{J}_{\mathcal{S}}$ occurs if the Jensen channel $\mathcal{H}_{\mathcal{S}}$ is in outage w.r.t. the rate $r(\mathcal{S}) \log \text{SNR}$, where $r(\mathcal{S}) = \sum_{u \in \mathcal{S}} r_u$, i.e., $\mathcal{J}_{\mathcal{S}} \triangleq \{J_{\mathcal{S}} < r(\mathcal{S}) \log \text{SNR}\}$. From (9) we can conclude that, obviously, $\mathbb{P}(\mathcal{J}_{\mathcal{S}}) \leq \mathbb{P}(\mathcal{O}_{\mathcal{S}})$.

We shall next characterize the Jensen outage probability analytically. Recalling (3), we start by writing $\mathcal{H}_{\mathcal{S}} = \mathcal{H}_w(\mathbf{R}^{T/2} \otimes \mathbf{I}_{\text{M}(\mathcal{S})})$, where $\mathbf{R} = \mathbf{R}_{\mathbb{H}}$, if $M_R \leq |\mathcal{S}|M_T$, and $\mathbf{R} = \mathbf{R}_{\mathbb{H}}^T$, if $M_R > |\mathcal{S}|M_T$, and $\mathcal{H}_w$ is the i.i.d. $\mathcal{CN}(0,1)$ matrix with the same dimensions as $\mathcal{H}_{\mathcal{S}}$ given by

$$\mathcal{H}_w = \begin{cases} [\mathbf{H}_{w,0} \ \mathbf{H}_{w,1} \ \ldots \ \mathbf{H}_{w,N-1}], & \text{if } M_R \leq |\mathcal{S}|M_T, \\ [\mathbf{H}_{w,0}^H \ \mathbf{H}_{w,1}^H \ \ldots \ \mathbf{H}_{w,N-1}^H], & \text{if } M_R > |\mathcal{S}|M_T. \end{cases}$$

Here, $\mathbf{H}_{w,n}$ denotes i.i.d. $\mathcal{CN}(0,1)$ matrices of dimension $M_R \times |\mathcal{S}|M_T$. Since $\mathcal{H}_w \mathbf{U} \sim \mathcal{H}_w$, for any unitary $\mathbf{U}$, and $\mathbf{R}_{\mathbb{H}}$ and $\mathbf{R}_{\mathbb{H}}^T$ have the same eigenvalues, we get $\mathcal{H}_{\mathcal{S}} \mathcal{H}_{\mathcal{S}}^H \sim \mathcal{H}_w (\mathbf{\Lambda} \otimes \mathbf{I}_{\text{M}(\mathcal{S})}) \mathcal{H}_w^H$, where $\mathbf{\Lambda} = \text{diag}\{\lambda_1(\mathbf{R}_{\mathbb{H}}), \lambda_2(\mathbf{R}_{\mathbb{H}}), \ldots, \lambda_\rho(\mathbf{R}_{\mathbb{H}}), 0, \ldots, 0\}$. With $\overline{\mathcal{H}}_w = \mathcal{H}_w([1:\text{m}(\mathcal{S})], [1:\rho\text{M}(\mathcal{S})])$, it was shown in [7] that $\mathbb{P}(\mathcal{J}_{\mathcal{S}})$ is nothing but the outage probability of an effective MIMO channel with $\rho\text{M}(\mathcal{S})$ transmit and $\text{m}(\mathcal{S})$ receive antennas and satisfies

$$\mathbb{P}(\mathcal{J}_{\mathcal{S}}) \doteq \mathbb{P}\left(\log \det\left(\mathbf{I} + \text{SNR}\, \overline{\mathcal{H}}_w \overline{\mathcal{H}}_w^H\right) < r(\mathcal{S}) \log \text{SNR}\right)$$
$$\doteq \text{SNR}^{-d_{\mathcal{S}}(r(\mathcal{S}))} \quad (11)$$

where we infer from the results in [1] that $d_{\mathcal{S}}(r)$ is the piecewise linear function connecting the points $(r, d_{\mathcal{S}}(r))$ for $r = 0, 1, \ldots, \text{m}(\mathcal{S})$, with

$$d_{\mathcal{S}}(r) = (\text{m}(\mathcal{S}) - r)(\rho\text{M}(\mathcal{S}) - r). \quad (12)$$

Since, as already noted, $\mathbb{P}(\mathcal{O}_{\mathcal{S}}) \geq \mathbb{P}(\mathcal{J}_{\mathcal{S}})$, it follows from (11) that

$$\mathbb{P}(\mathcal{O}_{\mathcal{S}}) \dotgeq \text{SNR}^{-d_{\mathcal{S}}(r(\mathcal{S}))}. \quad (13)$$

We shall see below that (13) is a key ingredient for establishing the optimal DM tradeoff.

*B. Error event analysis*

Following [2], [10], we decompose the total error probability into $2^U - 1$ disjoint error events according to

$$P_e(\mathcal{C}_r) = \sum_{\mathcal{S} \subseteq \mathcal{U}} \mathbb{P}(\mathcal{E}_{\mathcal{S}}) \quad (14)$$

where the $\mathcal{S}$-error event $\mathcal{E}_{\mathcal{S}}$ corresponds to all the users in $\mathcal{S}$ being decoded incorrectly and the remaining users being decoded correctly. More precisely, we have

$$\mathcal{E}_{\mathcal{S}} \triangleq \left\{(\hat{\mathbf{X}}_u \neq \mathbf{X}_u, \forall u \in \mathcal{S}) \wedge (\hat{\mathbf{X}}_u = \mathbf{X}_u, \forall u \in \bar{\mathcal{S}})\right\} \quad (15)$$

where $\mathbf{X}_u$ and $\hat{\mathbf{X}}_u$ are, respectively, the transmitted and ML-decoded codewords corresponding to user $u$. The following result establishes a DM tradeoff optimal code design criterion for a specific error event $\mathcal{E}_{\mathcal{S}}$.

*Theorem 1:* For every $u \in \mathcal{S}$, let $\mathcal{C}_{r_u}$ have block length $N \geq \rho|\mathcal{S}|M_T$, and set $\lambda_n = \lambda_n\big((\mathbf{R}_{\mathbb{H}}^T \odot (\sum_{u \in \mathcal{S}} \mathbf{E}_u^H \mathbf{E}_u))\big)$ for $n = 1, 2, \ldots, \rho|\mathcal{S}|M_T$, where $\mathbf{E}_u = \mathbf{X}_u - \mathbf{X}'_u$ and $\mathbf{X}_u, \mathbf{X}'_u \in \mathcal{C}_{r_u}(\text{SNR})$. Furthermore, define

$$\Lambda_{\text{m}(\mathcal{S})}^{\rho|\mathcal{S}|M_T}(\text{SNR}) \triangleq \min_{\substack{\mathbf{E}_u = \mathbf{X}_u - \mathbf{X}'_u, \forall u \in \mathcal{S} \\ \mathbf{X}_u, \mathbf{X}'_u \in \mathcal{C}_{r_u}(\text{SNR})}} \prod_{k=1}^{\text{m}(\mathcal{S})} \lambda_k. \quad (16)$$

If there exists an $\epsilon > 0$ such that

$$\Lambda_{\text{m}(\mathcal{S})}^{\rho|\mathcal{S}|M_T}(\text{SNR}) \dotgeq \text{SNR}^{-(r(\mathcal{S}) - \epsilon)}, \quad (17)$$

then, under ML decoding, $\mathbb{P}(\mathcal{E}_{\mathcal{S}}) \dotleq \text{SNR}^{-d_{\mathcal{S}}(r(\mathcal{S}))}$.

*Proof:* We start by deriving an upper bound on the average (w.r.t. the random channel) pairwise error probability (PEP) of an $\mathcal{S}$-error event. Let the codewords of $\mathcal{C}_r(\text{SNR})$ be given by $\mathbf{X} = [\mathbf{X}_1^T \ \mathbf{X}_2^T \ \ldots \ \mathbf{X}_U^T]^T$. Based on (15), we note that $\mathbf{E}_u = \mathbf{X}_u - \mathbf{X}'_u$ is nonzero for $u \in \mathcal{S}$ and $\mathbf{E}_u = \mathbf{0}$ for $u \in \bar{\mathcal{S}}$. Assuming, without loss of generality, that $\mathcal{S} = \{1, 2, \ldots, |\mathcal{S}|\}$, the probability of the ML decoder mistakenly deciding in favor of the codeword $\mathbf{X}'$ when $\mathbf{X}$ was actually transmitted can be upper bounded in terms of $\mathbf{X} - \mathbf{X}' = [\mathbf{E}_1^T \ \mathbf{E}_2^T \ \ldots \ \mathbf{E}_{|\mathcal{S}|}^T \ \mathbf{0} \ \ldots \ \mathbf{0}]^T$ as

$$\mathbb{P}(\mathbf{X} \to \mathbf{X}')$$
$$\leq \mathbb{E}_{\{\mathbf{H}_{\mathcal{S},n}\}}\left\{\exp\left(-\frac{\text{SNR}}{4M_T} \sum_{n=0}^{N-1} \text{Tr}\left(\mathbf{H}_{\mathcal{S},n} \mathbf{e}_n \mathbf{e}_n^H \mathbf{H}_{\mathcal{S},n}^H\right)\right)\right\} \quad (18)$$

where $\text{Tr}\left(\mathbf{H}_{\mathcal{S},n} \mathbf{e}_n \mathbf{e}_n^H \mathbf{H}_{\mathcal{S},n}^H\right) = \left\|\sum_{u \in \mathcal{S}} \mathbf{H}_{u,n} \mathbf{e}_{u,n}\right\|^2$ with $\mathbf{H}_{\mathcal{S},n}$ defined in (2) and $\mathbf{e}_n = [\mathbf{e}_{u_1,n}^T \ \mathbf{e}_{u_2,n}^T \ \cdots \ \mathbf{e}_{u_{|\mathcal{S}|},n}^T]^T$, where $\mathbf{e}_{u,n} = \mathbf{x}_{u,n} - \mathbf{x}'_{u,n}$. Defining $\mathbf{H}_{\mathcal{S}} = [\mathbf{H}_{\mathcal{S},0} \ \mathbf{H}_{\mathcal{S},1} \ \cdots \ \mathbf{H}_{\mathcal{S},N-1}]$, we get from (18)

$$\mathbb{P}(\mathbf{X} \to \mathbf{X}')$$
$$\leq \mathbb{E}_{\mathbf{H}_{\mathcal{S}}}\left\{\exp\left(-\frac{\text{SNR}}{4M_T}\text{Tr}\left(\mathbf{H}_{\mathcal{S}}\, \text{diag}\{\mathbf{e}_n \mathbf{e}_n^H\}_{n=0}^{N-1}\, \mathbf{H}_{\mathcal{S}}^H\right)\right)\right\}$$
$$= \mathbb{E}_{\mathbf{H}_w}\left\{\exp\left(-\frac{\text{SNR}}{4M_T}\text{Tr}\left(\mathbf{H}_w \boldsymbol{\Upsilon}\boldsymbol{\Upsilon}^H \mathbf{H}_w^H\right)\right)\right\} \quad (19)$$

where we used $\mathbf{H}_{\mathcal{S}} = \mathbf{H}_w(\mathbf{R}_{\mathbb{H}}^{T/2} \otimes \mathbf{I}_{|\mathcal{S}|M_T})$ with $\mathbf{H}_w$ an $M_R \times N|\mathcal{S}|M_T$ matrix with i.i.d. $\mathcal{CN}(0,1)$ entries and

$$\boldsymbol{\Upsilon} = (\mathbf{R}_{\mathbb{H}}^{T/2} \otimes \mathbf{I}_{|\mathcal{S}|M_T})\, \text{diag}\{\mathbf{e}_n\}_{n=0}^{N-1}. \quad (20)$$

Noting that $\boldsymbol{\Upsilon}^H \boldsymbol{\Upsilon} = \mathbf{R}_{\mathbb{H}}^T \odot (\sum_{u \in \mathcal{S}} \mathbf{E}_u^H \mathbf{E}_u)$ and using the fact that the nonzero eigenvalues of $\boldsymbol{\Upsilon}\boldsymbol{\Upsilon}^H$ in (19) equal the nonzero eigenvalues of $\boldsymbol{\Upsilon}^H \boldsymbol{\Upsilon}$, it follows, by assumption, that $\boldsymbol{\Upsilon}\boldsymbol{\Upsilon}^H$ has precisely $\rho|\mathcal{S}|M_T$ nonzero eigenvalues. The remainder of the proof proceeds along the lines of the proof of Theorem 1 in [13][3]. In particular, we split and subsequently bound the $\mathcal{S}$-error

---

[3]For the point-to-point case, the criterion in [13, Theorem 1] requires the $\text{m} = \min(M_T, M_R)$ smallest eigenvalues of the effective codeword difference matrix to satisfy $\prod_{k=1}^{\text{m}} \lambda_k \dotgeq \text{SNR}^{-(r-\epsilon)}$, whereas Theorem 1 in [7] requires $\lambda_{\min}^{\text{m}} \dotgeq \text{SNR}^{-(r-\epsilon)}$. It can readily be seen that the latter condition implies the former and, hence, the criterion in [13] provides a relaxed optimality condition.

probability as

$$\begin{aligned}\mathbb{P}\left(\mathcal{E}_{\mathcal{S}}\right) &= \mathbb{P}\left(\mathcal{E}_{\mathcal{S}}, \mathcal{J}_{\mathcal{S}}\right) + \mathbb{P}\left(\mathcal{E}_{\mathcal{S}}, \bar{\mathcal{J}}_{\mathcal{S}}\right) \\ &= \mathbb{P}\left(\mathcal{J}_{\mathcal{S}}\right)\underbrace{\mathbb{P}\left(\mathcal{E}_{\mathcal{S}}|\mathcal{J}_{\mathcal{S}}\right)}_{\leq 1} + \underbrace{\mathbb{P}\left(\bar{\mathcal{J}}_{\mathcal{S}}\right)}_{\leq 1}\mathbb{P}\left(\mathcal{E}_{\mathcal{S}}|\bar{\mathcal{J}}_{\mathcal{S}}\right) \\ &\leq \mathbb{P}\left(\mathcal{J}_{\mathcal{S}}\right) + \mathbb{P}\left(\mathcal{E}_{\mathcal{S}}|\bar{\mathcal{J}}_{\mathcal{S}}\right).\end{aligned} \qquad (21)$$

As detailed in [13], the code design criterion (17) yields the following upper bound on the second term in (21):

$$\mathbb{P}\left(\mathcal{E}_{\mathcal{S}}|\bar{\mathcal{J}}_{\mathcal{S}}\right) \leq \mathsf{SNR}^{Nr(\mathcal{S})} \exp\left(-\frac{\mathsf{SNR}^{\epsilon/\mathrm{m}(\mathcal{S})}}{4\mathrm{M_T}}\right). \qquad (22)$$

In contrast to the Jensen outage probability which satisfies $\mathbb{P}(\mathcal{J}_{\mathcal{S}}) \doteq \mathsf{SNR}^{-d_{\mathcal{S}}(r(\mathcal{S}))}$, (22) decays exponentially in SNR. Hence, upon inserting (22) into (21), we get $\mathbb{P}(\mathcal{E}_{\mathcal{S}}) \dotle \mathbb{P}(\mathcal{J}_{\mathcal{S}})$, and can therefore conclude that $\mathbb{P}(\mathcal{E}_{\mathcal{S}}) \dotle \mathsf{SNR}^{-d_{\mathcal{S}}(r(\mathcal{S}))}$. ■

In summary, for every $\mathcal{E}_{\mathcal{S}}$, we have a sufficient condition on $\{\mathcal{C}_{r_u} : u \in \mathcal{S}\}$ for $\mathbb{P}(\mathcal{E}_{\mathcal{S}})$ to be exponentially upper bounded by $\mathbb{P}(\mathcal{J}_{\mathcal{S}})$. Based on this result, we shall next establish the optimal DM tradeoff for the MAC and provide corresponding design criteria on the family $\mathcal{C}_{r}$.

*C. Optimal code design*

We start by noting that (5) implies $\mathbb{P}(\mathcal{O}) \geq \mathbb{P}(\mathcal{O}_{\mathcal{S}})$ for any $\mathcal{S} \subseteq \mathcal{U}$, which combined with (13) gives rise to $2^U - 1$ lower bounds on $\mathbb{P}(\mathcal{O})$. For a given multiplexing rate tuple $r$, the tightest lower bound (exponentially in SNR) corresponds to the set $\mathcal{S}$ that yields the smallest SNR exponent $d_{\mathcal{S}}(r(\mathcal{S}))$. More precisely, the tightest lower bound is characterized by

$$\mathbb{P}(\mathcal{O}) \dotge \mathsf{SNR}^{-d_{\mathcal{S}^\star}(r(\mathcal{S}^\star))} \qquad (23)$$

where the dominant outage event corresponds to the set

$$\mathcal{S}^\star = \arg\min_{\mathcal{S} \subseteq \mathcal{U}} d_{\mathcal{S}}(r(\mathcal{S})). \qquad (24)$$

Next, we show that, for any multiplexing rate tuple, the total error probability $P_e(\mathcal{C}_r)$ can be made exponentially equal to the lower bound in (23) by appropriate design of the users' codebooks. As a direct consequence thereof, using $P_e(\mathcal{C}_r) \dotge \mathbb{P}(\mathcal{O})$ [2, Lemma 7] and (23), we then obtain that $d_{\mathcal{S}^\star}(r(\mathcal{S}^\star))$ constitutes the optimal DM tradeoff of the selective-fading MIMO MAC.

*Theorem 2:* The optimal DM tradeoff of the selective-fading MIMO MAC in (1) is given by $d^\star(r) = d_{\mathcal{S}^\star}(r(\mathcal{S}^\star))$, that is

$$d^\star(r) = (\mathrm{m}(\mathcal{S}^\star) - r(\mathcal{S}^\star))(\rho\mathrm{M}(\mathcal{S}^\star) - r(\mathcal{S}^\star)). \qquad (25)$$

Moreover, if the family of codes $\mathcal{C}_r$ satisfies (17) for every $\mathcal{S} \subseteq \mathcal{U}$, then

$$d(\mathcal{C}_r) = d^\star(r). \qquad (26)$$

*Proof:* Inserting the upper bound (21) into (14), we get

$$\begin{aligned}P_e(\mathcal{C}_r) &\leq \sum_{\mathcal{S} \subseteq \mathcal{U}} \left(\mathbb{P}(\mathcal{J}_{\mathcal{S}}) + \mathbb{P}(\mathcal{E}_{\mathcal{S}}|\bar{\mathcal{J}}_{\mathcal{S}})\right) \\ &\dotle \sum_{\mathcal{S} \subseteq \mathcal{U}} \mathbb{P}(\mathcal{J}_{\mathcal{S}}) \qquad (27) \\ &\doteq \mathsf{SNR}^{-d_{\mathcal{S}^\star}(r(\mathcal{S}^\star))} \qquad (28)\end{aligned}$$

where (27) is a consequence of the assumption that $\mathcal{C}_r$ satisfies (17) for every $\mathcal{S} \subseteq \mathcal{U}$ and (28) follows from (11) together with the definition (24). With $P_e(\mathcal{C}_r) \dotge \mathbb{P}(\mathcal{O})$ [2, Lemma 7], combining (23) and (28) yields

$$P_e(\mathcal{C}_r) \doteq \mathbb{P}(\mathcal{O}) \doteq \mathsf{SNR}^{-d_{\mathcal{S}^\star}(r(\mathcal{S}^\star))}. \qquad (29)$$

Since, by definition, $d(\mathcal{C}_r) \leq d^\star(r)$, using (8), we can conclude from (29) that $d(\mathcal{C}_r) = d^\star(r) = d_{\mathcal{S}^\star}(r(\mathcal{S}^\star))$. ■

As a consequence of Theorem 2, the maximum achievable diversity order is determined by the error event that dominates the total error probability exponentially in SNR. To see this, let $\mathcal{E}_{\mathcal{S}'}$ denote the $\mathcal{S}$-error event that dominates the overall error probability so that, based on (14), $P_e(\mathcal{C}_r) \doteq \mathbb{P}(\mathcal{E}_{\mathcal{S}'})$. By (29), we necessarily have $d_{\mathcal{S}^\star}(r(\mathcal{S}^\star)) = d(\mathcal{E}_{\mathcal{S}'})$, where

$$d(\mathcal{E}_{\mathcal{S}'}) = -\lim_{\mathsf{SNR}\to\infty}\frac{\log\mathbb{P}(\mathcal{E}_{\mathcal{S}'})}{\log\mathsf{SNR}}.$$

Since $\mathcal{C}_r$ satisfies (17), Theorem 1 yields $d(\mathcal{E}_{\mathcal{S}'}) \geq d_{\mathcal{S}'}(r(\mathcal{S}'))$ and, hence, we get $d_{\mathcal{S}^\star}(r(\mathcal{S}^\star)) \geq d_{\mathcal{S}'}(r(\mathcal{S}'))$. However, by the definition of $\mathcal{S}^\star$, we also have $d_{\mathcal{S}^\star}(r(\mathcal{S}^\star)) \leq d_{\mathcal{S}'}(r(\mathcal{S}'))$ which implies $d_{\mathcal{S}^\star}(r(\mathcal{S}^\star)) = d_{\mathcal{S}'}(r(\mathcal{S}'))$. Thus, $P_e(\mathcal{C}_r) \doteq \mathbb{P}(\mathcal{E}_{\mathcal{S}^\star})$, which is to say that the optimal DM tradeoff is given by the SNR exponent corresponding to the dominant error event.

*Example:* We assume $\mathrm{M_T} = 3$, $\mathrm{M_R} = 4$, and $\rho = 2$. For $U = 2$, the $2^2 - 1 = 3$ possible error events are denoted by $\mathcal{E}_1$ (user 1 only is in error), $\mathcal{E}_2$ (user 2 only is in error) and $\mathcal{E}_3$ (both users are in error). The SNR exponents of the corresponding error probabilities are obtained from (12) as

$$\begin{aligned}d_u(r_u) &= (3 - r_u)(8 - r_u), \quad u = 1, 2, \\ d_3(r_1 + r_2) &= \left(4 - (r_1 + r_2)\right)\left(12 - (r_1 + r_2)\right).\end{aligned} \qquad (30)$$

Based on (30), we can now explicitly determine the dominant error event for every multiplexing rate tuple $r = (r_1, r_2)$. In Figure 1, we plot the rate regions dominated by the different error events. Note that the SNR exponent of the error probability is zero whenever $r_1 > 3$, $r_2 > 3$ or $r_1 + r_2 > 4$. In the rate region dominated by $\mathcal{E}_1$, we have $d_1(r_1) < d_2(r_2)$ and $d_1(r_1) < d_3(r_1+r_2)$, implying that the SNR exponent of the total error probability equals $d_1(r_1)$, i.e., the SNR exponent that would be obtained in a point-to-point selective-fading MIMO channel with $\mathrm{M_T} = 3$, $\mathrm{M_R} = 4$, and $\rho = 2$. The same reasoning applies to the rate region dominated by $\mathcal{E}_2$ and, hence, we can conclude that, in the sense of the DM tradeoff, the performance in regions $\mathcal{E}_1$ and $\mathcal{E}_2$ is not affected by the presence of the second user. In contrast, in the area dominated by $\mathcal{E}_3$, we have $d_3(r_1 + r_2) < d_u(r_u)$, $u = 1, 2$, which is to say that the multiuser interference does have an impact on the DM tradeoff and reduces the diversity gain that would be obtained if only one user were present.

V. AN OPTIMAL CODE FOR THE FLAT-FADING CASE

Satisfying the code design criterion (17) for every $\mathcal{S} \subseteq \mathcal{U}$ is non-trivial and systematic procedures for designing DM tradeoff optimal codes are an important open problem. In this section, we show that the algebraic code construction proposed recently in [9] for flat-fading MACs with single-antenna users satisfies (17) for every $\mathcal{S} \subseteq \mathcal{U}$ and any multiplexing rate tuple[4].

---
[4]In [9], the DM tradeoff optimality of the proposed code is shown for $r_1 = r_2$.

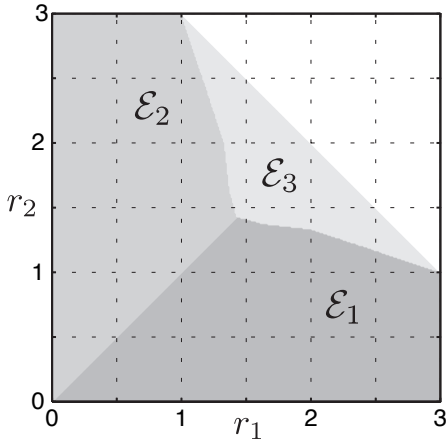

Fig. 1. Dominant error event regions for the two-user MIMO MAC with $M_T = 3$, $M_R = 4$, and $\rho = 2$.

We start by briefly reviewing the code construction described in [9] for a system with $M_T = 1$, $M_R = 2$, $U = 2$, $N = 2$, and $\rho = 1$. For each user $u$, let $\mathcal{A}_u$ denote a QAM constellation with $2^{R_u(\mathsf{SNR})}$ points carved from $\mathbb{Z}[i] = \{k + il : k, l \in \mathbb{Z}\}$, where $i = \sqrt{-1}$. The proposed code spans two slots so that the vector of information symbols corresponding to user $u$ is given by $\mathbf{s}_u = [s_{u,0} \; s_{u,1}]$, where $s_{u,0}, s_{u,1} \in \mathcal{A}_u$. Using the unitary transformation matrix $\mathbf{U}$ underlying the Golden Code [5], the $1 \times 2$ codeword $\mathbf{X}_u$ is obtained as

$$\mathbf{X}_u^T = \mathbf{U}\,\mathbf{s}_u^T = \begin{bmatrix} x_u \\ \sigma(x_u) \end{bmatrix} \text{ with } \mathbf{U} = \frac{1}{\sqrt{5}}\begin{bmatrix} \alpha & \alpha\varphi \\ \bar{\alpha} & \bar{\alpha}\bar{\varphi} \end{bmatrix} \quad (31)$$

where $\varphi = \frac{1+\sqrt{5}}{2}$ denotes the Golden number with corresponding conjugate $\bar{\varphi} = \frac{1-\sqrt{5}}{2}$, $\alpha = 1 + i - i\varphi$ and $\bar{\alpha} = 1 + i - i\bar{\varphi}$. Here, $\sigma$ denotes the generator of the Galois group of the quadratic extension $\mathbb{Q}(i, \sqrt{5})$ over $\mathbb{Q}(i) = \{k + il : k, l \in \mathbb{Q}\}$ given by

$$\sigma : \begin{array}{rcl} \mathbb{Q}(i, \sqrt{5}) & \to & \mathbb{Q}(i, \sqrt{5}) \\ a + b\sqrt{5} & \mapsto & a - b\sqrt{5}. \end{array} \quad (32)$$

Moreover, one of the users, say user 2, multiplies the symbol corresponding to the first slot by a constant $\gamma \in \mathbb{Q}(i)$, resulting in the overall $2 \times 2$ codeword

$$\mathbf{X} = \begin{bmatrix} x_1 & \sigma(x_1) \\ \gamma x_2 & \sigma(x_2) \end{bmatrix}. \quad (33)$$

As shown in [9], for any $\gamma \neq \pm 1$ and any two $\mathbf{X}, \mathbf{X}'$ according to (33), it holds that $\det(\mathbf{\Delta}) \neq 0$, where $\mathbf{\Delta} = \mathbf{X} - \mathbf{X}'$. For the so-defined construction we have the following result.

*Theorem 3:* For any multiplexing rate tuple $\mathbf{r}$, the algebraic code construction in [9] satisfies (17) for any $\mathcal{S} \subseteq \mathcal{U}$.

*Proof:* We start by assuming that at any given SNR, user $u$ carves out $2^{R_u(\mathsf{SNR}) - \epsilon \log \mathsf{SNR}}$ points from $\mathbb{Z}[i]$ for some $\epsilon > 0$, i.e., $|\mathcal{A}_u| = \mathsf{SNR}^{r_u - \epsilon}$. In order to satisfy the power constraint (7), we scale $\mathcal{A}_u$ by $\mathsf{SNR}^{-(r_u - \epsilon)/2}$ so that, due to the linearity (over $\mathbb{C}$) of the transformation in (31), the codeword corresponding to user $u$ is given by $\mathsf{SNR}^{-(r_u - \epsilon)/2}\,\mathbf{X}_u$. From

(33) and the linearity of the mapping $\sigma$ over $\mathbb{Q}(i, \sqrt{5})$, the codeword difference matrix is obtained as

$$\mathbf{E} = \begin{bmatrix} \mathsf{SNR}^{-(r_1-\epsilon)/2} e_1 & \mathsf{SNR}^{-(r_1-\epsilon)/2}\sigma(e_1) \\ \mathsf{SNR}^{-(r_2-\epsilon)/2}\gamma e_2 & \mathsf{SNR}^{-(r_2-\epsilon)/2}\sigma(e_2) \end{bmatrix} \quad (34)$$

where $e_u = x_u - x'_u$, $u = 1, 2$. Next, we note that in the flat-fading case $\mathbf{R}_{\mathbb{H}}^T \odot (\mathbf{E}^H \mathbf{E}) = \mathbf{E}^H \mathbf{E}$. In particular, considering user 1, i.e., $\mathcal{S} = \{1\}$, we have $|\mathcal{S}| = 1$ and $\mathrm{m}(\mathcal{S}) = 1$ so that, from (16), we obtain $\Lambda_1^1(\mathsf{SNR}) = \mathsf{SNR}^{-(r_1-\epsilon)} \min_{e_1}(|e_1|^2 + |\sigma(e_1)|^2)$. Letting $\mathbf{X}_1^T = \mathbf{U}\mathbf{s}_1^T$ and $(\mathbf{X}'_1)^T = \mathbf{U}(\mathbf{s}'_1)^T$ and since $\mathbf{U}$ is unitary, we get $(|e_1|^2 + |\sigma(e_1)|^2) = \|\mathbf{s}_1 - \mathbf{s}'_1\|^2 \geq 2 d_{\min}^2$, where $d_{\min}$ is the (nonzero) minimum distance in $\mathcal{A}_1$. We therefore conclude that $\Lambda_1^1(\mathsf{SNR}) \doteq \mathsf{SNR}^{-(r_1-\epsilon)}$. For user 2, a similar argument shows that $\Lambda_1^1(\mathsf{SNR}) \doteq \mathsf{SNR}^{-(r_2-\epsilon)}$ and, hence, the construction satisfies the criteria arising from (17) for $\mathcal{S} = \{1\}$ and $\mathcal{S} = \{2\}$. For $\mathcal{S} = \{1, 2\}$, note that $|\mathcal{S}| = 2$ and $\mathrm{m}(\mathcal{S}) = 2$ so that $\Lambda_2^2(\mathsf{SNR}) = \min_{\mathbf{E}} |\det(\mathbf{E})|^2$. From (34), we get $|\det(\mathbf{E})|^2 = \mathsf{SNR}^{-(r_1+r_2-2\epsilon)} |\det(\mathbf{\Delta})|^2$. Recalling that for any $\gamma \neq \pm 1$, $\det(\mathbf{\Delta})$ is nonzero and independent of SNR, it follows that $|\det(\mathbf{E})|^2 \doteq \mathsf{SNR}^{-(r_1+r_2-2\epsilon)}$ and, consequently, we obtain $\Lambda_2^2(\mathsf{SNR}) \doteq \mathsf{SNR}^{-(r_1+r_2-2\epsilon)}$, from which we can conclude that (17) is also satisfied for $\mathcal{S} = \{1, 2\}$. The proof is concluded by taking $\epsilon$ to be arbitrarily close to zero, implying that both users operate arbitrarily close to their target multiplexing rates. ∎

ACKNOWLEDGMENT

The first author thanks Cemal Akçaba for stimulating discussions on the proof of Theorem 2.